\documentclass[twocolumn,aps,showpacs,floatfix,superscriptaddress]{revtex4}
\usepackage{amsmath,amssymb,eucal,graphicx,color}

\begin{document}

\title{Approximating Spectral Impact of Structural Perturbations in Large Networks}

\author{Attilio Milanese}
{\thanks{Current address: Technology Department - MSC -MDA, CERN CH-1211, Geneva (Switzerland); E-mail: \texttt{amilanes@cern.ch}}
\affiliation{Department of Mechanical \& Aeronautical Engineering, Clarkson University, Potsdam, NY 13699-5725}

\author{Jie Sun}
{\thanks{Current address: Department of Physics \& Astronomy, Northwestern University, Evanston, IL 60208-3112;\\ 
Email: sunj@northwestern.edu}
\affiliation{Department of Mathematics \& Computer Science,  Clarkson University, Potsdam, NY 13699-5815}

\author{Takashi Nishikawa}
\email{tnishika@clarkson.edu}
\affiliation{Department of Mathematics \& Computer Science,  Clarkson University, Potsdam, NY 13699-5815}

\begin{abstract}
Determining the effect of structural perturbations on the eigenvalue spectra of networks is an important problem because the spectra characterize not only their topological structures, but also their dynamical behavior, such as synchronization and cascading processes on networks.
Here we develop a theory for estimating the change of the largest eigenvalue of the adjacency matrix or the extreme eigenvalues of the graph Laplacian when small but \emph{arbitrary} set of links are added or removed from the network.
We demonstrate the effectiveness of our approximation schemes using both real and artificial networks, showing in particular that we can accurately obtain the spectral ranking of small subgraphs.
We also propose a local iterative scheme which computes the relative ranking of a subgraph using only the connectivity information of its neighbors within a few links.
Our results may not only contribute to our theoretical understanding of dynamical processes on networks, but also lead to practical applications in ranking subgraphs of real complex networks.
\end{abstract}


\maketitle


\section{Introduction}
The theory and application of complex networks have been a popular and exciting research topic since the seminal work \cite{watts1998collective, Barabasi:1999kc} appeared at the end of last century. The study of complex networks interests a variety of fields, such as mathematics, physics, computer science, sociology, and biology, to list a few. 
(For excellent reviews, see for example Refs.~\cite{Albert:2002gd, NEWMAN_SIAM03, Dorogovtsev:2008ly}.)
One of the most important problems in complex networks, both from a theoretical and applicative viewpoint, is that of measuring centrality. Various measures of centrality have been proposed based on different quantitative properties of the underlying network. Examples are degree centrality, shortest path and random walk betweenness, clustering coefficient and eigenvector component \cite{NEWMAN_SIAM03, ESTRADA_PRE05}. 
Common applications of such measures is the ranking of vertices of the network, in particular in the context of web search engines \cite{KLEINBERG_ACM99, LANGVILLE_SIAM05}. Many of these centrality measures are connected with the spectral radius of the adjacency matrix of the graph; furthermore, the spectral radius itself is crucial for a class of dynamical processes on networks \cite{MAC_SIAM, Juan-G.-Restrepo:2006ak, Restrepo:2006fk, Dorogovtsev:2008ly}. 

A fundamental question in the study of networks is then how the spectral radius and other invariant network statistics change under structural perturbations, such as the removal or addition of a few links, or the modification of their strength.
This question is particularly relevant in the context of evolving networks, since the stepwise changes in such networks are typically small, and developing efficient algorithms to track the corresponding changes in network statistics is crucial in understanding the dynamics of network evolution~\cite{SUN_PRE09}.
The effect of structural perturbations on the spectral radius as well as other network statistics has significant consequences for the security and robustness of networked systems under component failures and intentional attacks.

Let $G = (V,E,Q)$ be a weighted graph, where $V$ is the set of vertices, $E$ the set of directed edges and $Q$ the set of weights on the edges. 
For convenience, 
we label the vertices in $V$ with integers $1, \ldots, n$. 
Let $A \in \mathbb{R}^{n\times n}$ be the adjacency matrix of the graph, where its entry $a_{ij}$ is defined as the weight on the edge going from node $i$ to node $j$. If $A$ is non-negative and irreducible, then the Perron-Frobenius theorem~\cite{LANCASTER,MAC_SIAM} can be used to show that its largest eigenvalue $\lambda$ is non-degenerate and positive, and there exist positive left and right eigenvectors associated with it. In this case, denote the normalized left and right eigenvectors by $u$ and $v$, respectively ($u, v \in\mathbb{R}^{n}$), so that
\begin{equation} \label{eq:lreigs}
	u^T A = \lambda u^T, \;\;	
	Av = \lambda v, \;\; \left\|u\right\|_2 = \left\|v\right\|_2 = 1.
\end{equation}
The condition of non-negativity is equivalent to requiring 
non-negative weights on edges; 
on the other hand, the condition of irreducibility corresponds to the graph being strongly connected. In this paper, the graphs under consideration all share these properties.

In Ref.~\cite{Restrepo:2006fk} the {\it dynamical importance} $I_{ij}$ of an edge from $i$ to $j$ was defined as the amount of relative decrease the removal of this edge causes on $\lambda$, i.e., if after removing such an edge $\lambda$ becomes $\lambda - \Delta\lambda_{ij}$, then $I_{ij} \equiv \Delta\lambda_{ij}/\lambda$. Hence the dynamical importance of an edge quantitatively captures the effect that the removal of such an edge has on the largest eigenvalue of the graph adjacency matrix. An analogous definition was introduced for the dynamical importance of vertices, where $I_k$ is the importance of node $k$. Approximations of these dynamical importances, based on perturbation techniques, are given in Ref.~\cite{Restrepo:2006fk}; in particular, $I_{ij}$ and $I_k$ can be approximated, respectively, by
\begin{equation}\label{eq:approx_I_ij_1}
	\bar{I}_{ij} = \frac{a_{ij} u_i v_j}{\lambda u^{T} v}
\end{equation}
and
\begin{equation} \label{eq:approx_I_k_1}
	\bar{I}_k = \frac{u_k v_k}{u^{T} v},
\end{equation}
where $u_i$ and $v_i$ denote the $i$-th component of the vectors $u$ and $v$, respectively.

In this paper, the notion of dynamical importance is extended to 
measure the {\em spectral impact} \footnote{This name is chosen to reflect the fact that the underlying measure is the spectral radius and that we consider not only removal but also addition of links.} of a structural perturbation in which an \emph{arbitrary} set of links are removed or added, or their weights are modified.
We define the spectral impact (SI) as
\begin{equation}\label{def:SI}
	I_{B} \equiv \frac{\left.\lambda\right|_{A+B} - \left.\lambda\right|_A}{\left.\lambda\right|_A},
\end{equation}
where $A$ denotes the adjacency matrix of the original graph, $\left.\lambda\right|_A$ is the corresponding spectral radius, 
while $B$ is the adjacency matrix of the graph consisting of all the $n$ nodes and the links that are removed, added, or modified. 
Here a positive value of $I_B$ corresponds to an increase in $\lambda$, and a negative value to a decrease.
Note that in the case of removing a subgraph of $A$, the entries of $B$ will be non-positive.
In this framework, the edge and vertex importance can be treated as special cases. 
To enable efficient estimations of the spectral impact, formulae based on the first and improved first order approximations are presented and discussed in detail. These formulae lead to the observation that the degree centrality, as a local measure, can be viewed as a first order approximation to the eigenvector centrality.

The rest of the paper is organized as follows. Section~II reports first and second order perturbation results to approximate the SI using global information on the graph; in particular, the special cases of changing the weight on an edge and removing a node are addressed. In Section~III, applications to synthetic and real-world graphs are presented. Section~IV deals with estimating the SI using local information on the network. While the majority of the results concerns the spectral radius of the adjacency matrix of the graph, Section~V deals with an extension of our method to the graph Laplacian. 


\section{Approximating Spectral Impact using Global Information}

\subsection{Perturbation results}

Suppose a graph is modified so that its adjacency matrix $A$ becomes $A + \epsilon C$. Let $\lambda$ be the largest eigenvalue of $A$, and $\lambda+\Delta\lambda$ the largest eigenvalue of $A + \epsilon C$.
The change $\Delta\lambda$ can be estimated, when $\epsilon\ll 1$, using a first-order perturbation result \cite{WILKINSON, GOLUB, LANCASTER, DEMMEL}, as
\begin{equation} \label{eq:approx_lambda_1old}
	\Delta\lambda \approx \Delta_1 = \epsilon \left.\lambda'\right|_A = \frac{u^T \epsilon C v}{u^T v},
\end{equation}
where 
$\left.\lambda'\right|_A$ denotes the derivative of $\lambda+\Delta\lambda$ as a function of $\epsilon$, evaluated at $\epsilon=0$.
Equation~\eqref{eq:approx_lambda_1old} shows that---at the first order---the change in the largest eigenvalue is obtained from 
the derivative $\left.\lambda'\right|_A$, which depends on both $A$ and $C$ but can be computed using the vectors $u$ and $v$, and the matrix $C$. If $A$ is symmetric, $u = v$ and Eq.~\eqref{eq:approx_lambda_1old} is of the Rayleigh quotient form~\cite{DEMMEL}.

For more accurate approximation, we can use a second-order perturbation result:
\begin{equation}\label{eq:2nd_order_approx}
	\Delta\lambda \approx \Delta_1 + \Delta_2 = \epsilon \left.\lambda'\right|_A + \frac{\epsilon^2}{2} \left. \lambda''\right|_A .
\end{equation}
Computing the second derivative term using known perturbation results would require knowledge of all the eigenvalues and eigenvectors of $A$~\cite{WILKINSON}. This is impractical. Hence, a further approximation is introduced, as
\begin{equation} \label{eq:approx_lambda''}
	\left. \lambda''\right|_A \approx \frac{\left.\lambda'\right|_{A + \epsilon C} - \left.\lambda'\right|_A}{\epsilon}.
\end{equation}
The term $\left.\lambda'\right|_{A + \epsilon C}$ takes into account the change in $u$ and $v$, to $u + \Delta u$ and $v + \Delta v$, respectively. It is proposed here to estimate $v + \Delta v$ by means of one iteration of the power method \footnote{Details on the power method are reported in Section IV. Here, the assumption is that $v$ and $v + \Delta v$ are close to being parallel, and hence a single iteration of the power method is satisfactory. The spectral gap of $A + \epsilon C$ is not critical for this approximation.}, starting from the unperturbed eigenvector $v$, as
\begin{align} 
	v + \Delta v & \approx \frac{(A + \epsilon C) v}{\left\| (A + \epsilon C) v \right\|_2} = 
		\frac{\lambda v + \epsilon C v}{\left\| \lambda v + \epsilon C v \right\|_2} = \nonumber \\
		& = \frac{\lambda v + \epsilon C v}{\sqrt{ \lambda^2 v^T v + 2 \epsilon \lambda v^T C v + \epsilon^2 v^T C^T C v }}
			\nonumber \\
	  & \approx \frac{\lambda v + \epsilon C v}{\lambda},
\end{align}
where $\epsilon$ terms in the denominator have been neglected. Thus
\begin{equation} \label{eq:approx_delta_v}
	\Delta v \approx \frac{\epsilon}{\lambda} C v.
\end{equation}
Similarly, it can be found that
\begin{equation} \label{eq:approx_delta_u}
	\Delta u \approx \frac{\epsilon}{\lambda} C^T u.
\end{equation}
Therefore, the derivative $\left.\lambda'\right|_{A + \epsilon C}$ can be approximated as
\begin{align} 
	\left.\lambda'\right|_{A + \epsilon C} 
	& = \frac{\left(u + \Delta u\right)^T C \left(v + \Delta v \right)}
		{\left(u + \Delta u\right)^T \left(v + \Delta v \right)} \approx \nonumber \\
	& \approx \frac{u^T C v}{u^T v} + \frac{u^T C \Delta v}{u^T v} + \frac{\Delta u^T C v}{u^T v} \approx \nonumber \\
	& \approx \frac{u^T C v}{u^T v} + \frac{2 \epsilon}{\lambda} \frac{u^T C C v}{u^T v},
\end{align}
having neglected terms containing $\Delta u$ and $\Delta v$ in the denominator, and the product $\Delta u^T C \Delta v$ in the numerator.
Thus, an improved approximation
for $\Delta\lambda$ is given by
\begin{equation} \label{eq:approx_lambda_2old}
	\Delta\lambda \approx \Delta_1 + \Delta_2 \approx \frac{u^T \epsilon C v}{u^T v} + \frac{1}{\lambda} \frac{u^T \epsilon^2 C^2 v}{u^T v}.
\end{equation}

The above formulae take advantage of the fact that in many situations the eigenvector does not change much by the perturbation, and thus a few steps (one or two) of the power method already give a very accurate approximation, as we shall see in the next section. 
This, however, is not the case when the dominant eigenvalue $\lambda$ is nearly degenerate, in which case the convergence rate of the power method is $\bigl\lvert\frac{\lambda_2}{\lambda}\bigr\rvert\approx{1}$, where $\lambda_2$ is the second largest eigenvalue in absolute value.
More precisely, the accuracy of our approximation is related to the spectral gaps through the perturbation formula, which, for undirected networks, takes the form~\footnote{Here the formula is presented for undirected networks for convenience, and the argument holds similarly in general cases.}
\begin{equation}\nonumber
\Delta\lambda = \epsilon{v^T C v} + \epsilon^{2}\sum_{i=2}^{n}\dfrac{(v^{(i)}Cv)^2}{\lambda-\lambda_i} + o(\epsilon^2),
\end{equation}
where $\lambda_i$ and $v^{(i)}$ ($i\geq{2}$) are the 
non-dominant eigenvalues and corresponding eigenvectors of $A$. When $\lambda_1\approx\lambda_2$, the second term becomes large, indicating that $\lambda$ is highly sensitive to perturbations.

\subsection{Adding or removing an arbitrary subgraph}
Most of the interesting problems in the context of networks involve discrete changes, where $A$ becomes $A + B$, and the nonzero entries of $B$ are finite and usually of the same order of magnitude as the nonzero entries of $A$.
If, however, the modifications are limited to a small number of links in a large network, then $\left\|B\right\|_2 \ll \left\|A\right\|_2$, and the perturbation results \eqref{eq:approx_lambda_1old} and \eqref{eq:2nd_order_approx} would be valid with $\epsilon C$ replaced by $B$.
The other approximations we have made are also likely to be valid, and this is supported by the fact that Eq.~\eqref{eq:approx_lambda_2old} improves significantly over Eq.~\eqref{eq:approx_lambda_1old} for the example networks discussed in Section \ref{application}.
Equations~\eqref{eq:approx_lambda_1old} and \eqref{eq:approx_lambda_2old} then become
\begin{align} 
\Delta\lambda &\approx \Delta_1 = \frac{u^T B v}{u^T v} \label{eq:approx_lambda_1_B} \\
\Delta\lambda &\approx \Delta_1 + \Delta_2 \approx \frac{u^T B v}{u^T v} + \frac{1}{\lambda} \frac{u^T B^2 v}{u^T v} \label{eq:approx_lambda_2_B}
\end{align}
and our approximation schemes for the spectral impact $I_B$ for small but finite modifications are
\begin{align} 
\hat{I}_B &= \frac{1}{\lambda} \frac{u^T B v}{u^T v} \label{eq:approx_lambda_1}\\
\hat{\hat{I}}_B &= \frac{1}{\lambda} \frac{u^T B (v+ Bv/\lambda) }{u^T v}. \label{eq:approx_lambda_2}
\end{align}
Equation~\eqref{eq:approx_lambda_1} is linear in $B$, and therefore the change can be decomposed into the sum of elementary changes, as $B = \sum_i B_i$, where $B_i$ can represent, for example, a modification of a single edge. The corresponding first order approximation for the SI is obtained from the individual contributions, as $\hat{I}_B = \sum_i \hat{I}_{B_i}$. On the other hand, Eq.~\eqref{eq:approx_lambda_2} has a linear and a quadratic dependence on $B$ and linear superposition cannot in general be used. However, if all the products of elementary changes $B_i B_j$ are zero matrices
(for example, if $B_i$'s represent the disconnected components of the subgraph), 
then $\hat{\hat{I}}_B = \sum_i \hat{\hat{I}}_{B_i}$.

When the changes are limited to a small number of links, we have a clear computational advantage of having highly sparse matrix $B$. Also, including changes in the eigenvector does not imply much computational burden. In fact, the extra computational cost of using Eq.~\eqref{eq:approx_lambda_2_B} instead of Eq.~\eqref{eq:approx_lambda_1_B} amounts to computing the extra vector $u^T B$ and projecting it onto $B v$ \footnote{The vector $B v$ is considered to be available from the first order approximation. Alternatively, one could proceed by computing $u^T B$ at the first order and then $B v$: the increase in the computational cost is, in general, the same.}.

\subsection{Modifying weight on single edge}

Adding weight $b_{ij}$ to 
directional
edge $(i,j)$ of a graph \footnote{As particular case, taking $b_{ij} = -a_{ij}$ amounts to deleting the edge $(i,j)$. On the other hand, if $a_{ij} = 0$, then adding $b_{ij}$ means introducing a directed connection with a certain weight.} corresponds to changing the adjacency matrix from $A$ to $A + B_{ij}$, where $B_{ij}$ is an $n\times n$ matrix containing all zeros except for the $(i,j)$ entry $b_{ij}$. The spectral impact $I_{B_{ij}}$ can be approximated, using Eq.~\eqref{eq:approx_lambda_1}, as
\begin{equation}\label{eq:remove_edge}
	\hat{I}_{B_{ij}} = \frac{b_{ij} u_i v_j}{\lambda u^{T} v}.
\end{equation}
This is equivalent to Eq.~\eqref{eq:approx_I_ij_1}, in the case $b_{ij}=-a_{ij}$.

The second order approximation~\eqref{eq:approx_lambda_2} introduces a correction only if a modification on a self-loop is introduced, as $(B^2)_{ij}$ 
is zero otherwise. Indeed, $I_{B_{ij}}$ can be approximated at the second order as
\begin{equation}
	\hat{\hat I}_{B_{ij}} = \frac{b_{ij} u_i v_j}{\lambda u^{T} v} ( 1 + \delta_{ij}\frac{b_{ii}}{\lambda} ),
\end{equation}
where $\delta_{ij}$ is the Kronecker delta.

If the change is introduced 
{\it bidirectionally}, 
that is, $b_{ij}$ is added to the weight of edge $(i,j)$, and $b_{ji}$ to edge $(j,i)$, then the first and second order approximations to the SI become 
\begin{equation} \label{eq:approx_lambda_1_symm}
	\hat{I}_{B_{ij,ji}} = \frac{b_{ij} u_i v_j + b_{ji} u_j v_i}{\lambda u^{T} v}
\end{equation}
and
\begin{equation} \label{eq:approx_lambda_2_symm}
	\hat{\hat I}_{B_{ij,ji}} = \frac{b_{ij} u_i v_j + b_{ji} u_j v_i}{\lambda u^{T} v} + 
		\frac{b_{ij} b_{ji} \left( u_i v_i + u_j v_j \right)}{\lambda^2 u^{T} v}.
\end{equation}
In this case Eq.~\eqref{eq:approx_lambda_2_symm} can have nonzero correction term for modification of weights on an edge other than a self loop.  For undirected networks, modification must be bidirectional and symmetric ($b_{ij}=b_{ji})$, and the formulae become
\begin{equation} \label{eq:approx_lambda_1_undirected}
	\hat{I}_{B_{ij}} = \frac{2 b_{ij} v_i v_j}{\lambda}
\end{equation}
and
\begin{equation} \label{eq:approx_lambda_2_undirected}
	\hat{\hat I}_{B_{ij}} = \frac{2 b_{ij} v_i v_j}{\lambda} + 
		\frac{b_{ij}^2 \left( v_i^2 + v_j^2 \right)}{\lambda^2}.
\end{equation}
Note that Eq.~\eqref{eq:approx_lambda_2_undirected} always introduces a non-negative correction to the first order approximation for modifications to undirected networks.

\subsection{Removing a node}

Removing a node $k$ in a graph corresponds to removing all edges touching it. In this case,
the entries of $B$ are ${b}_{ij} = -a_{ij}(\delta_{ik} + \delta_{jk} - \delta_{ik}\delta_{jk})$ and
the first order approximation reads
\begin{equation} 
	\hat{I}_k = \left( -2 + \frac{a_{kk}}{\lambda} \right) \frac{u_k v_k}{u^T v},
\end{equation}
while the second order one yields
\begin{equation} 
	\hat{\hat I}_k = \left(-1 + \frac{a_{kk}}{\lambda} - \frac{a_{kk}^2}{\lambda^2} + 
		\frac{1}{\lambda^2} \sum_{i=1}^n a_{ik} a_{ki} \right) \frac{u_k v_k}{u^T v}.
\end{equation}

In the case that there is no self-loop connecting node $k$ with itself, the above formulae simplify to
\begin{equation} 
	\hat{I}_k = -2 \frac{u_k v_k}{u^T v}
\end{equation}
and
\begin{equation} 
	\hat{\hat I}_k = \left(-1 + \frac{1}{\lambda^2} \sum_{i=1}^n a_{ik} a_{ki} \right) \frac{u_k v_k}{u^T v}.
\end{equation}
The summation term in the last equation represents an improvement over the previous result, Eq.~\eqref{eq:approx_I_k_1}.

If, furthermore, the network is undirected and unweighted, first and second order approximations become
\begin{equation} \label{eq:approx_lambda_1_node_last}
	\hat{I}_k = -2 v_k^2
\end{equation}
and
\begin{equation} \label{eq:approx_lambda_2_node_last}
	\hat{\hat I}_k = \left(-1 + \frac{d^\text{out}_k}{\lambda^2} \right) v_k^2,
\end{equation}
where $d^\text{out}_k \equiv \sum_{i=1}^{n}a_{ki}$ is the {\it out-degree} of vertex $k$.

The last two equations show well that, in the case of removing a node, first and second order approximations yield rather different results, with the estimate from Eq.~\eqref{eq:approx_lambda_1_node_last} more than double that from Eq.~\eqref{eq:approx_lambda_2_node_last}. 
This difference comes from the fact that, in the case of removing node $k$, the $k$-th component of the new dominant eigenvector becomes zero, regardless of its previous value, and thus the change in $u$ and $v$ is not negligible.

\section{Applications\label{application}}

In this section, the accuracy of 
various approximations 
are assessed using both synthetic and real-world graphs. 
The true dominant eigenvalue and corresponding eigenvector are estimated in double precision using the MATLAB function \texttt{eigs} \cite{MATLAB_2006} and compared to the estimates obtained by implementing our approximation formulae.

As an example of artificial networks, the Erd\"os-R\'enyi random graph~\cite{BOLLOBAS} is used, with $n = 1000$ nodes and the probability of connection $p=0.01$. The particular realization used is labelled $\mathcal{G}_1$ for convenience and has 5004 undirected links, without self-loops. The largest degree is $d_{max} = 20$, the minimum is $d_{min} = 2$, and the average is $d_{mean} = 10.01$. The largest three eigenvalues (in magnitude) of the corresponding adjacency matrix are computed to be $\lambda = 11.0741$, $\lambda_2 = -6.53518$, and $\lambda_3 = 6.50196$. The components of the eigenvector $v$ are shown in increasing order in Fig.~\ref{fig:results_G1}(a).
We also analyze three real-world networks, which we refer to as $\mathcal{G}_2$, $\mathcal{G}_3$, and $\mathcal{G}_4$ for convenience. The basic properties of these networks are reported in Table~\ref{tab:ExamplesOfRealWorldNetworks}, along with the pertinent references. $\mathcal{G}_2$ is a biological example, $\mathcal{G}_3$ is a social interaction network, while $\mathcal{G}_4$ can be regarded as an instance having both an engineering and social character. 

\begin{table}
\centering
\caption{Examples of real world networks.}
  \begin{tabular}{l}
  \\ \hline
			$\mathcal{G}_2$, yeast protein interaction network 		
					\cite{bu2003topological} \\
			2361 vertices, 13828 edges \\
			$d_{min} = 1$, $d_{mean} = 5.86$, $d_{max} = 65$ \\
			$\lambda = 19.4861$, $\lambda_2 = 16.1340$, $\lambda_3 = 14.3339$ 
	\\ \hline
			$\mathcal{G}_3$, network of e-mail interchanges
					\cite{guimera2003self} \\
			1133 vertices, 5451 edges \\
			$d_{min} = 1$, $d_{mean} = 9.62$, $d_{max} = 71$ \\
			$\lambda = 41.4940$, $\lambda_2 = 33.9272$, $\lambda_3 = 30.0687$ 
	\\ \hline
			$\mathcal{G}_4$, US power grid\footnote{The data was retrieved from P. Tsaparas' webpage on data sets and codes for complex networks, at \url{http://www.cs.helsinki.fi/u/tsaparas/MACN2006/data-code.html}.}
			4941 vertices, 6594 edges \\
			$d_{min} = 1$, $d_{mean} = 2.67$, $d_{max} = 19$ \\
			$\lambda = 7.4831$, $\lambda_2 = 6.6092$, $\lambda_3 = 5.5728$
	\\ \hline
  \end{tabular}
\label{tab:ExamplesOfRealWorldNetworks}
\end{table}

The impact on $\lambda$ of the removal of various subgraphs of $\mathcal{G}_1$ is analyzed next: 
the results from the first and second order approximations are plotted versus the actual SI. In Fig.~\ref{fig:results_G1}(b), the removal of edges is considered; the maximum reduction in $\lambda$ is about 0.1\% and it is satisfactorily predicted by both the first and second order formulae: the former has a tendency of overestimating the change, while the latter is more accurate. As expected, there are edges that have a greater impact on $\lambda$ than others. In particular, the impact of edge $(i,j)$ is dictated by the components $v_i$ and $v_j$ of $v$, as given in 
Eqs.~\eqref{eq:approx_lambda_1_undirected} and \eqref{eq:approx_lambda_2_undirected}.
The effect of removing a simple \textit{motif} is shown in Fig.~\ref{fig:results_G1}(c), where the $171$ triangles occurring in $\mathcal{G}_1$ are individually removed. The observed $\Delta\lambda$ is higher in this case, though the approximations are still satisfactory. Figure~\ref{fig:results_G1}(d) investigates the removal of nodes, comparing Eqs.~\eqref{eq:approx_lambda_1_node_last}, \eqref{eq:approx_lambda_2_node_last} and \eqref{eq:approx_I_k_1}: while the first order approximation is off by a factor of about $2$, the second order formula with the $d_k/\lambda^2$ term is the most accurate. 
Note that for both the first and second order estimates the plots are almost monotonic, indicating that the relative ranking of nodes, edges, and triangles defined by the SI are accurately estimated by these formulae.
Indeed, for a randomly chosen pair of nodes or edges, the estimate is correct with probability close to one (Table~\ref{tab:rank_preserving}).

\begin{table*}
  \caption{Predicting the relative ranking of edges and nodes based on SI.  The numbers indicate the fraction of all possible pairs of edges or nodes for which the relative ranking is correctly predicted by each approximation scheme.  
  The local methods are based on the eigenvector components estimated by the normalized degree (LM1), the normalized sum of the neighbors' degree (LM2), and the normalized sum of the degrees of the neighbors' neighbors (LM3).}
  \begin{tabular}{@{\extracolsep{5pt}}c@{\hspace{15pt}}cccccp{1pt}@{\hspace{20pt}}cc@{\hspace{15pt}}ccc}
  \hline
   & \multicolumn{5}{c}{Ranking of edges} & & \multicolumn{5}{c}{Ranking of nodes}\\[5pt]
  & & & \multicolumn{3}{c}{Local Methods} & & & &  \multicolumn{3}{c}{Local Methods}\\
  Networks   & Eq.~\eqref{eq:approx_lambda_1_B}  & Eq.~\eqref{eq:approx_lambda_2_B} & LM1 & LM2 & LM3 & & Eq.~\eqref{eq:approx_lambda_1_B}  & Eq.~\eqref{eq:approx_lambda_2_B}  & LM1 & LM2 & LM3\\
  \hline
   $\mathcal{G}_1$    &   99.34\%  &  99.95\% & 90.20\% & 96.63\% & 98.04\% & &  99.59\%  &  99.94\%    &  93.79\%  &  96.81\%  &  97.86\% \\
   $\mathcal{G}_2$    &   98.46\%  &  99.95\% & 82.31\% & 88.01\% & 90.13\% & &  99.85\% &   99.96\%    &  84.30\%  &  88.81\%   & 91.52\% \\
   $\mathcal{G}_3$    &   98.96\%  &  99.97\% & 85.08\% & 91.40\% & 93.85\% & & 99.86\%  &  99.95\%   &   87.65\%  &  91.88\%  &  94.64\% \\
   $\mathcal{G}_4$    &   96.86\%  &  96.51\% & 83.68\% & 91.07\% & 93.60\% & &  96.59\%  & 95.92\%     &  85.91\%  &  82.40\%  &  81.99\% \\
   \hline
  \end{tabular}
\label{tab:rank_preserving}
\end{table*}

\begin{figure}
\includegraphics[width=3.5in]{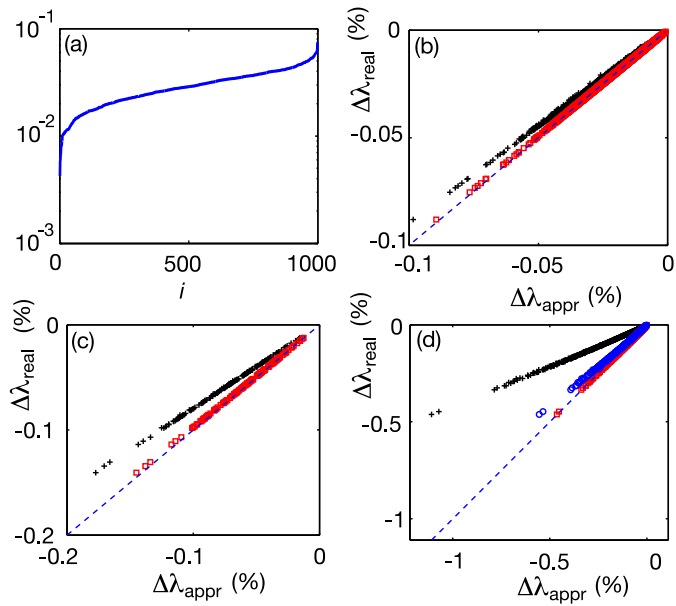}
\caption{(Color online) Results for the Erd\"os-R\'enyi network $\mathcal{G}_1$.  (a) Eigenvector components sorted in the increasing order.  The true SI is plotted against its approximation (both in percentage) for the removal of (b) edges, (c) triangles, and (d) nodes.   The black plus symbols correspond to the first order approximation~\eqref{eq:approx_lambda_1} and the red (light gray) squares to the improved approximation~\eqref{eq:approx_lambda_2}.  In (d) the blue (dark gray) circles correspond to the approximation in Ref.~\cite{Restrepo:2006fk}.}
\label{fig:results_G1}
\end{figure}

Figure~\ref{fig:results_G2_G3_G4} shows the eigenvector components, the real SI, and the approximated SI for removing an edge and removing a node from the real-world networks $\mathcal{G}_2$, $\mathcal{G}_3$, and $\mathcal{G}_4$. The eigenvector components are shown in the order of increasing magnitude; in several cases, the smallest components appear to be rather small and fall below the axis limit on the figure. The presence of components in the dominant eigenvector spanning several orders of magnitude amounts to large discrepancies in the importance of edges. When the removal of edges is analyzed, both the first and second order approximations for the SI are satisfactory. On the other hand, if removal of nodes is considered, the second order formula of Eq.~\eqref{eq:approx_lambda_2_node_last}, containing a correction for the degree of the node, yield results more accurate than Eq.~\eqref{eq:approx_I_k_1}. In this case, the SI is as large as -7\%.
The relative ranking based on SI is also accurately predicted for these real-world networks, as shown in Table~\ref{tab:rank_preserving}.
Note that increasing the order of approximation improves the accuracy, except for the case of $\mathcal{G}_4$.
For node removals in $\mathcal{G}_4$, the rank prediction accuracy, although high in general, falls slightly with the increasing order of approximation.
This is due to relatively large fluctuation of the approximation error among different nodes [Fig.~\ref{fig:results_G2_G3_G4}(i)], which is a likely consequence of heterogeneity and hierarchical nature of the power-grid network topology.

Since the spectral gap $\lambda-\lambda_2$ mainly determines the sensitivity of $\lambda$ to structural perturbations, it is directly related to the accuracy of our approximation formulae.
Indeed, the larger the spectral gap, the better the approximation will generally be.
Networks with large spectral gaps are known to be homogeneous and well-connected, avoiding structural bottlenecks~\cite{PhysRevE.75.016103}, and can also be characterized by having large expansion constant~\cite{Alon:1985fj,Hoory:2006lr}.

\begin{figure*}
\includegraphics{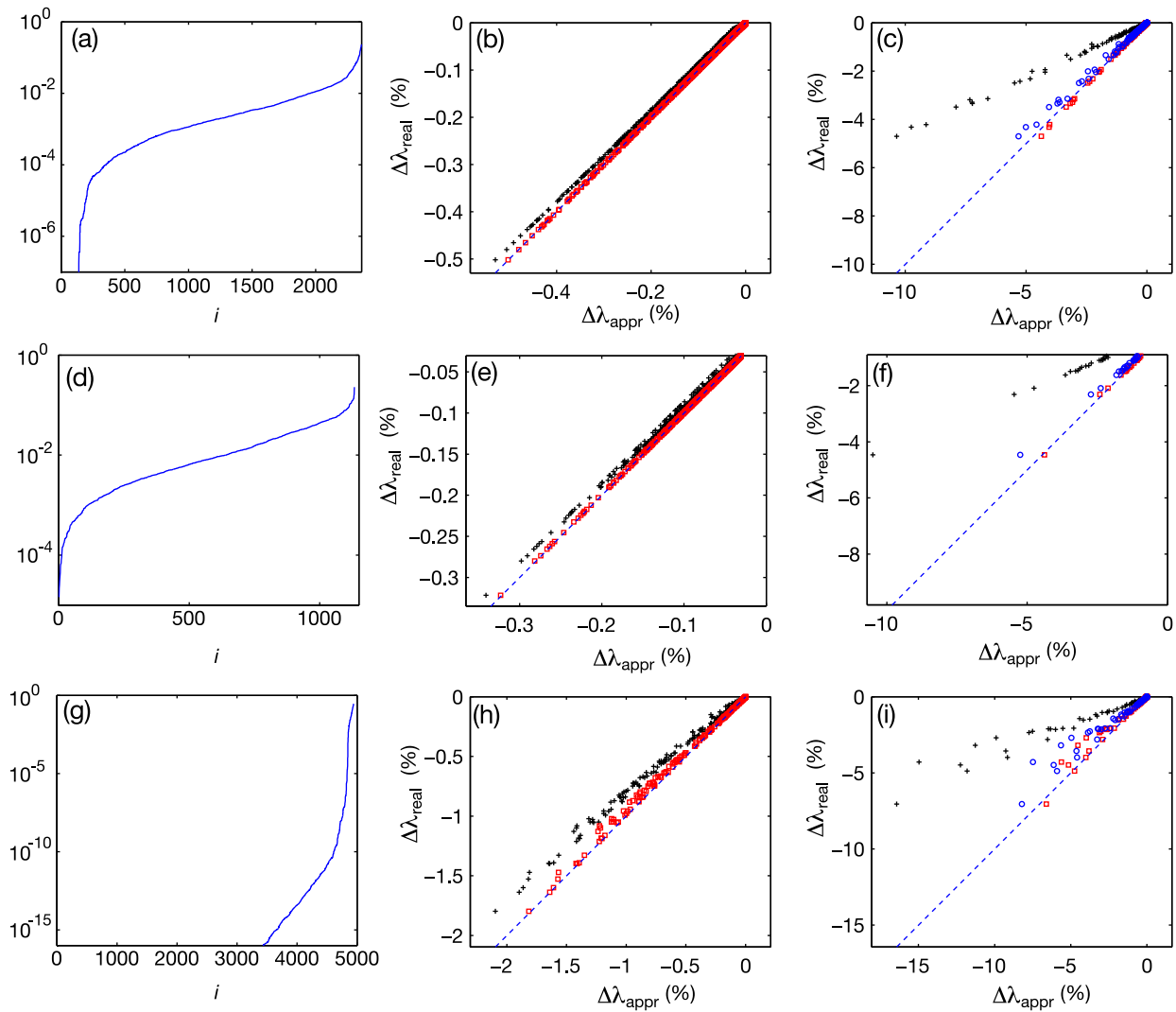}
\caption{(Color online) Results for the three examples of real-world networks $\mathcal{G}_2$ (top row), $\mathcal{G}_3$ (middle row) and $\mathcal{G}_4$ (bottom row) in Table~\ref{tab:ExamplesOfRealWorldNetworks}.  The left column [(a), (d), and (g)] shows the eigenvector components sorted in the increasing order.   The middle column [(b), (e), and (h)] shows the true SI vs approximated SI for the removal of edges, while the right column [(c), (f), and (i)] is for the removal of nodes.  The meaning of the symbols is the same as in Fig.~\ref{fig:results_G1}.}
\label{fig:results_G2_G3_G4}
\end{figure*}


\section{Approximating SI using Local Information}

In Section II, formulae for approximating the SI were given, based on Eq.~\eqref{eq:approx_lambda_1}.
However, the knowledge of both the largest eigenvalue $\lambda$ and its corresponding left and right eigenvectors $u$ and $v$ is required. Such information is in some situations impractical or even impossible to obtain, since it is equivalent to solving an eigenvalue problem for the adjacency matrix $A$, which requires knowing all the entries of $A$.

For a given set of links to be modified, however, our approximation formulae only require --- aside from a normalization constant --- a few components of $u$ and $v$ corresponding to the nodes attached to these links.
These $u$ and $v$ components can be approximated by iterative methods.
When using the SI for ranking different subgraphs,
such as single edges or nodes, or pairs of edges or nodes, 
this approach can be useful because the normalization constant does not affect the ranking.

Assuming that the spectrum of $A$ satisfies $|\lambda| > |\lambda_{2}| \geq \ldots \geq |\lambda_{n}|$, one can adopt the power method~\cite{WILKINSON, GOLUB, DEMMEL} to solve for the dominant eigenvalue $\lambda$ and its corresponding left and right eigenvectors $u$ and $v$. The starting point of this method is a normalized vector $v^{(0)}$, that in this case can be taken as
\begin{equation}\label{eq:v0}
v^{(0)} = \frac{1}{\sqrt{n}} [1,1,\cdots,1]^{T}.
\end{equation}
Then, for $t = 1,2,\ldots$, until convergence, the following is iterated:
\begin{equation}
	y^{(t)} = A v^{(t-1)}, \;	\lambda^{(t)} = ||y^{(t)}||_2, \; v^{(t)} = y^{(t)} / ||y^{(t)}||_2.
\end{equation}
The convergence of both the eigenvalue and eigenvector is geometric, with rate $O (|\lambda_{2}/\lambda|^{t})$
\footnote{This convergence result is easy to obtain if $A$ is diagonalizable.  For non-diagonalizable cases, see Ref.~\cite{WILKINSON} for details.}.
This algorithm can be straightforwardly adjusted for the computation of the left eigenvector $u$, and the same convergence rates apply.
Although other iterative schemes for the computation of the dominant eigenvalue and the corresponding eigenvector are available, the power method is used here since its iterations directly highlight the local information of a graph. 

Indeed, the first iteration of the power method gives the degree, the number of connections that each node has, up to a normalization constant. 
The $k$th component of $u^{(1)}$ is proportional to the {\it in-degree} of node $k$, $d_k^\text{in} = \sum_{i=1}^n a_{ik}$, while the $k$th component of $v^{(1)}$ scales with the {\it out-degree} of the same node, $d_k^\text{out} = \sum_{i=1}^n a_{ki}$.
If the graph is undirected and unweighted, the $k$th component of both $u^{(1)}$ and $v^{(1)}$ are proportional to the number of direct neighbors that node $k$ has (counting itself, if a self-loop is in place). 

The second iteration of the power method
returns information about the number of connections that these neighbors have.
For an undirected unweighted graph, the $k$th component of $u^{(2)}$ and $v^{(2)}$ is proportional to the sum of the degrees of the direct neighbors (where the same node can be counted several times, and node $k$ itself is always included in the count). 
If the graph is directed or weighted, the proper weights need to be added and the directions of the connections considered.
Subsequent iterations provide better approximations of the $k$th eigenvector component, involving larger neighborhood of node $k$.

Combing these with Eq.~\eqref{eq:approx_lambda_1}, we obtain successive approximations to the SI.
For edge $(i,j)$ in an undirected unweighted network, for example, the first two iterations give
\begin{align}
\hat{I}^{(1)}_{B_{ij}} &\sim d_i^\text{in} d_j^\text{out} \label{1st_iteration}\\
\hat{I}^{(2)}_{B_{ij}} &\sim \left( \sum_{k=1}^n a_{ki} d_k^\text{in} \right) \left(  \sum_{k=1}^n a_{jk} d_k^\text{out} \right). \label{2nd_iteration}
\end{align}
Figure~\ref{fig:results_local_G1} shows the results of the local computation for the Erd\"os-R\'enyi graph $\mathcal{G}_1$.
Similar results can be obtained for the networks $\mathcal{G}_2$, $\mathcal{G}_3$, and $\mathcal{G}_4$, which allow accurate prediction of edge- and node-ranking, as shown in Table~\ref{tab:rank_preserving}.
Analogously to the effect of the order of approximation, increasing the number of iterations for the local method improves the accuracy for all cases, except for $\mathcal{G}_4$ (due to the same large fluctuation of the approximation error among different nodes).
The accuracy of this local approximation also depends on the ratio $|\lambda_2/\lambda|$, since it dictates the convergence rate of the method.

\begin{figure}
\vspace{0.1in}
\includegraphics{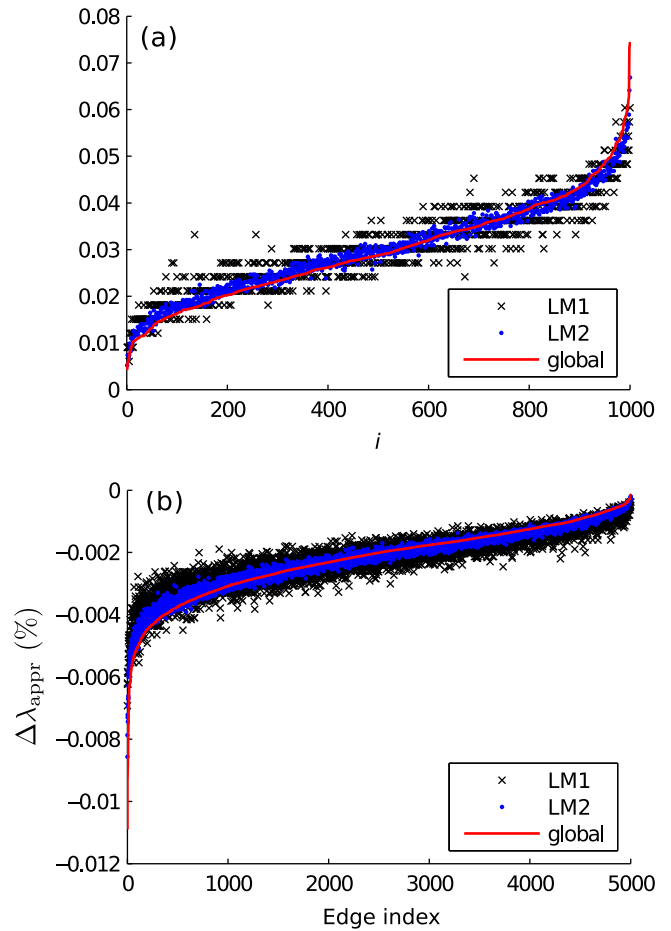}
\caption{(Color online) Convergence of local SI approximation for edge removal in the Erd\"os-R\'enyi network $\mathcal{G}_1$. 
(a) The eigenvector components sorted in the increasing order of their corresponding values computed globally [red (light gray) line].
LM1 denotes the approximation by the normalized degree (black cross symbols), while LM2 denotes the approximation by the normalized sum of the degrees of the neighbors [blue (dark gray) dots].  
(b) The SI for edge removal estimated using the approximations of the eigenvector in (a), showing quick convergence to the SI computed using the globally computed eigenvector.  The first order formula~\eqref{eq:approx_lambda_1_undirected} was used to estimate the SI.
}
\label{fig:results_local_G1}
\end{figure}

\section{Perturbation of the Graph Laplacian}

Perturbation results, based on either global or local information, can analogously be derived for eigenvalues and eigenvectors associated with the {\it graph Laplacian} $L$, defined as $L = D-A$, where $D$ is the diagonal matrix of node in-degrees. For simplicity, only connected and undirected graphs are considered, so that the associated graph Laplacians are positive-semidefinite matrices. The spectrum of $L$ is assumed to satisfy
\begin{equation}
	0 = \mu_{1} < \mu_{2} < \mu_{3} \leq ... \leq \mu_{n-1} < \mu_{n},
\end{equation}
so that $\mu_{2}$ and $\mu_{n}$ are non-degenerate.
The second smallest eigenvalue, $\mu_{2}$, and the largest one, $\mu_{n}$, are often used to characterize properties of the underlying graph. In particular, $\mu_{2}$ is known as the {\it algebraic connectivity} of the graph \cite{fiedler1989laplacian} and quantifies the connectedness of the graph ($\mu_2 = 0$ if the graph is disconnected).  
The algebraic connectivity, as well as the eigenratio $\mu_{2}/\mu_{n}$, are closely related to the stability of synchronized states in coupled dynamical systems~\cite{Pecora:1998zp,Sun:2009hc}.

Denote by $x$ and $y$ the normalized eigenvectors related to $\mu_{2}$ and $\mu_{n}$, respectively, 
\begin{equation}
	Lx = \mu_{2} x, \;\;	Ly = \mu_{n} y, \;\; \left\|x\right\|_2 = \left\|y\right\|_2 = 1.
\end{equation}
If an arbitrary subgraph (assumed to be symmetric and with no self loop) is added to or removed from the original graph, then the adjacency matrix changes from $A$ to $A + B$. The change in $\mu_{2}$ and $\mu_{n}$ can then be approximated, at the first order, by
\begin{equation}
	\Delta \mu_{2} \approx \sum_{i<j}{b_{ij}(x_{i}-x_{j})^{2}}
\end{equation}
and
\begin{equation}
	\Delta \mu_{n} \approx \sum_{i<j}{b_{ij}(y_{i}-y_{j})^{2}}.
\end{equation}
Combining the two equations above and neglecting terms containing more than one $\Delta$ (either $\Delta \mu_{2}$ or $\Delta \mu_{n}$) the change in the eigenratio $r = \mu_2/\mu_n$ is predicted by 
\begin{equation}
	\Delta r \approx \frac{1}{\mu_{n}^2}\sum_{i<j}{b_{ij}\big[\mu_{n}(x_{i}-x_{j})^{2} - \mu_{2}(y_{i}-y_{j})^{2}\big]}.
\end{equation}

Similarly to the formula for the dominant eigenvalue of $A$, the above equations can be adopted to develop strategies for targeting the network evolution towards some desired state, for example, to enhance (or weaken) the network synchronizability.  Our formalism can also be extended to other spectral quantities.


\section{Summary and Conclusions}

In this paper, we have introduced the concept of spectral impact of an arbitrary link modification in a network as the relative change in the largest eigenvalue of the adjacency matrix induced by the modification.  
Based on the standard perturbation method and an approximation for the second order term, we obtained an improved approximation formulae for the spectral impact that requires only the most dominant eigenvalue of the original network and its left and right eigenvectors.
Using the Erd\"os-R\'enyi random graph, as well as real-world examples of large complex networks from biological, social, and technological applications, we confirmed the accuracy of the formulae for the addition and/or removal of nodes, links, and triangles.  
We have also shown that the first iteration of a local approximation scheme based on the power method is equivalent to using the node (or subgraph) degree for ranking, and that further iterations quickly improve the accuracy by incorporating the connectivity structure of larger neighborhood of the node (or subgraph).
The analysis leading to the approximation formulae readily applies to the spectrum of other relevant matrices associated with the network, such as the Laplacian matrix treated briefly in this paper and the biased adjacency matrix studied in Ref.~\cite{ott:056111}.

Some problems on the approximation schemes still remain open.
How does the network topological structures, such as the small-world, scale-free, and modular structures, affect the accuracy of the approximations?
More generally, how does the robustness of the network with respect to its spectral properties depend on the network structure, and can it be used to classify networks, similarly to the existing spectral classification~\cite{PhysRevE.75.016103}?
How can we appropriately measure the ``smallness'' of perturbation $B$ to predict the accuracy?
It is also important to extend our method to the case of degenerate dominant eigenvalues, which may arise when the network evolved under constraints or under pressure to optimize its functions~\cite{Nishikawa:2006kx,Nishikawa:2006fk}.

Our results have several potential applications for large networks whose performance depends on their spectral properties.
The approximation schemes may be used in a damage control strategy for such networks, in which sudden structural damage that cannot be immediately fixed, such as the removal of multiple edges or nodes, is compensated by changes in other parts of the network (see Ref.~\cite{Motter:2008rm} for an example of such a compensatory perturbation in a different context). 
They may also be used to develop gradient-descent-like algorithms to solve the problem of designing networks that satisfy specific spectral (and thus dynamical) properties~\cite{HAGBERG_PRE06,Sun:2008tg,Hagberg:2008wd}, or to develop efficient updating schemes for spectrum-based statistics of evolving networks, similar to those for local statistics reported in Ref.~\cite{SUN_PRE09}.
The ranking of subgraphs, or motifs, according to the spectral impact in a given network and for a selected eigenvalue (not necessarily the largest) gives rise to interesting questions:  how does this ranking depend on the subgraph, the choice of the eigenvalue, and local and global properties of the network? Our improved formula reflects the fact that there is a nonlinear effect: the SI of the union of two subgraphs is not simply the sum of their individual SI. How does this nonlinear effect correlate with topological structures such as communities?
With the tools developed in this paper, we wish to tackle some of the above open problems and applications in our future work.


\acknowledgments 
A.M. gratefully acknowledges the partial support of the MAE Department through an instructor's assistantship. J.S. has been supported by the ARO grant 51950-MA. The authors would like to thank D.~ben-Avraham, A.~Alhakim, E.~M.~Bollt, and J.~D.~Skufca for numerous discussions.


\end{document}